# High-temperature stability of ambient-cured one-part alkali-activated materials incorporating graphene for thermal energy storage

Nghia P. Tran [1, 2, †], Tuan N. Nguyen [2, *, †], Jay R. Black [3, 4], Tuan D. Ngo [1,2, *]

[1] Building 4.0 CRC, Caulfield East, VIC 3145, Australia

[2] Department of Infrastructure Engineering, The University of Melbourne, VIC 3010, Australia

[3] School of Geography, Earth and Atmospheric Sciences, The University of Melbourne, VIC 3010, Australia

[4] Trace Analysis for Chemical, Earth and Environmental Sciences (TrACEES) Platform, The University of Melbourne, VIC 3010, Australia

**Graphical abstract**

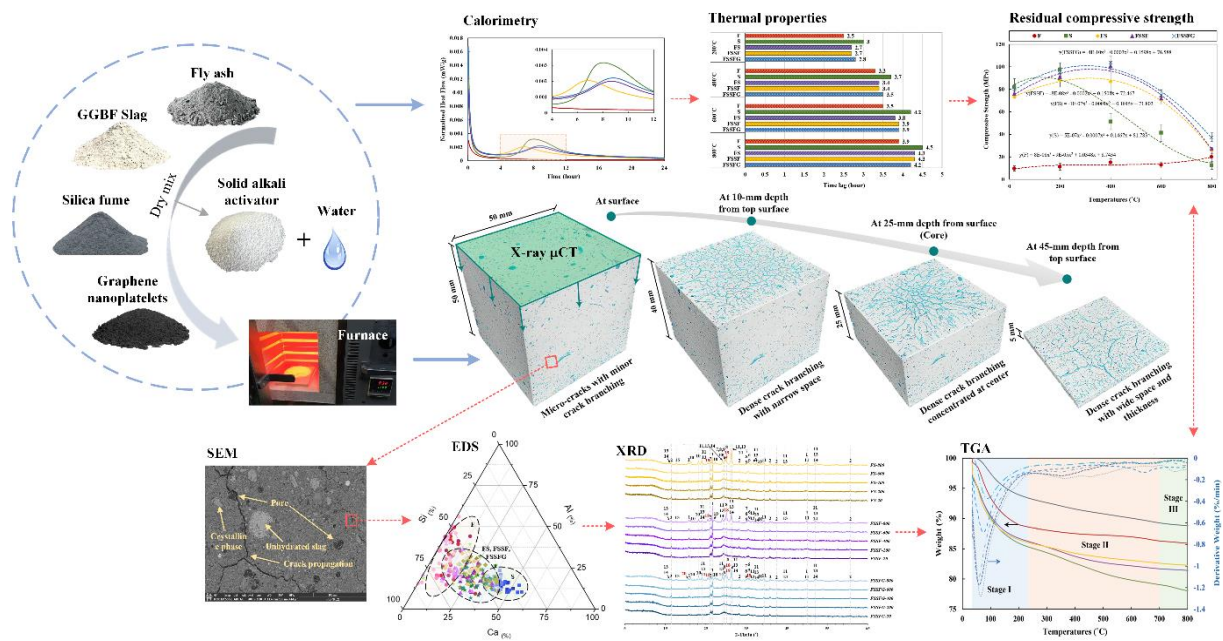

[*] Corresponding authors. E-mail address: tuan.nguyen@unimelb.edu.au (Tuan Nguyen); dtngo@unimelb.edu.au (Tuan Duc Ngo).

[†] These authors contributed equally to this work.

**Abstract**

In this research, the ambient-cured one-part alkali-activated material (AAM) containing graphene nanoplatelets (GNPs), fly ash, slag and silica fume has been investigated after high-temperature exposure to 200 – 800ºC. Their compressive strength, thermal properties, microstructure, pore structure were characterised through visual observation, isothermal calorimetry, TGA, XRD, SEM/EDS and X-ray mCT. The research findings indicated high strength characteristics of the developed AAM (~ 80 MPa) at ambient condition, which could further reach to approx. 100 MPa after being heated up to 400ºC. GNPs provided nucleation effects for promoting geopolymerisation and crystallisation. As observed from X-ray CT, a high extent of severe cracks initiated from the core and propagated towards the surface. From SEM/EDS analysis, high Na/Al and Na/Si ratios or low Si/Al and Ca/Si ratios highly correlated to thermal stability. Overall, the research outcomes implied the promising use of the nano-engineered AAMs for thermal energy storage (TES) at 400 – 600ºC.

# 1. Introduction

Global energy consumption has doubled that of the last four decades. With the ever-increasing trend in energy demand of society, global energy consumption amounted to around 440 EJ (i.e., ~122,000 TWh) in 2021 [1], and can be projected to triple by 2050 [2]. The heavy reliance on fossil energy sources (e.g., oil, gas, coal), nonetheless, associates with many pressing issues regarding greenhouse gas emissions, climate change and high energy costs. To resolve the current energy crisis and shift away from these polluted and expensive sources, accelerating the transition to renewable energy coupling with advanced energy storage technologies is the key. In this context, concentrated solar plants with integrated thermal energy storage (TES) systems offer a reliable and low-cost option to overcome the intermittent nature of renewable energy power. These advantages give solar power combined with TES technology an edge over wind and photovoltaic energy sources [3]. In TES systems, concrete is emerging as a solid storage medium for heat storage technology, which is highly competitive to the widely-used molten salt storage with several drawbacks as to high investment cost and high freezing points [4]. Extensive research has been conducted to demonstrate the thermo-mechanical performance and energy storage capacity of concrete for TES systems [5-12]. In real-world projects, concrete TES pilots have been successfully tested and deployed in large solar power plants, especially in the US, Spain, Germany and United Arab Emirates [13-17], where its feasibility has been proven. Upgrading the efficiency of these concrete TES systems will mainly rely on the enhanced thermal performance of core materials.

Sustainable and durable materials have gained extensive attention globally in a bid to alleviate the adverse impacts on the environment [18-21]. Among all, alkali-activated material (AAM) is a novel class of low-embodied carbon binders formed by aluminosilicate precursors (e.g., fly ash, slag, metakaolin, red mud, mine tailings) and alkaline activators (i.e., alkaline substances containing $OH^-$, $CO_3^{2-}$, $SiO_3^{2-}$ groups) [22-24]. With a similar strength to the traditional Ordinary Portland Cement (OPC) concrete, AAM emits 55 – 75% less $CO_2$ during its production [25]. The use of by-products waste as aluminosilicate precursors renders AAM sustainable, cost-effective and thermally stable [26-29]. More importantly, N-A-S-H or K-A-S-H polymeric chains in AAM withstand elevated temperatures (up to 800 °C) with little gel structural degradation, which obtain higher thermal stability than C-S-H phase in traditional cement-based materials [30]. The degree of geopolymerisation is accelerated corresponding to increasing exposure to elevated temperature [31, 32]. In literature, blended AAMs demonstrated a synergictic effect on facilitating the formation of temperature-stable crystalline phases (e.g., akermanite and gehlenite), thereby exhibiting enhanced thermal resilience at elevated temperatures [33]. In the study of Keane et al. [34], the developed waste-based AAM experienced minimal mass loss, less phase change and excellent thermal properties after several hundred thermal cycles (635 – 800°C). Rahjoo et al. [35] also tested the TES prototype using AAM as a solid storage medium and proved that the heat storage capacity of this AAM-based system (465 – 942 MJ/m$^3$) achieved 2 – 3.5 times higher than those of patented Heatcrete® (260 – 265 MJ/m$^3$) with OPC-based system [36], indicating the potential use of AAM for high-temperature TES in the future.

Two-dimensional (2D) carbon-based nanomaterials, such as graphene and its derivatives, have attracted the attention of researchers due to their superb thermo-electro-mechanical properties and high-temperature stability [37-40]. Acting as a catalyst, they accelerate the dissolution of amorphous phases, electron transmission as well as the transformation of $[SiO_2(OH)_2]^{2-}$ and $[Al(OH)_4]^-$ to the $[AlSiO_4]^-$ or $[AlSi_3O_8]^-$ structure, which promotes the growth of geopolymer gels and zeolite formation [41, 42]. The lower ionization potential of graphene-based materials can attract $Na^+/K^+$ from activator fluid to neutralise the adsorptive layer and induce particle coagulation, especially in high-alkaline solutions [43]. In addition, these carbon-based nanomaterials provide not only nanopore filling effects to densify the microstructural morphology [44, 45] but also cross-linking interaction with geopolymer chain to suppress crack propagation [46, 47]. Such mechanisms have proved the compatibility of graphene derivatives in geopolymer matrices and eventually result in the enhanced strength/toughness/durability of geopolymer composites [48-51]. However, their dosage is critical and should remain within an optimum range to ensure good dispersion as well as avert the disruption of agglomerated graphene to gel structure development during the geopolymerisation process [52].

Among graphene derivatives, graphene nanoplatelets (GNPs) have emerged as a competitive material due to their multi-layer honeycomb lattice geometry. GNPs is also a highly thermally conductive materials with thermal conductivity range from 6 W/mK (perpendicular to surface) to 3000 W/mK (parallel to surface) [53], and specific heat capacity of 643–2100 J/Kg°C [54]. Their high specific surface areas facilitate a stronger contact with geopolymeric binders [55], and provide additional nucleation sites for hydration and aluminosilicate reaction [56]. The increased thickness of GNPs also renders it less prone to agglomeration and entanglement [57], compared to monolayer graphene or carbon nanotubes (CNTs) [55]. As reported, a small concentration of GNPs (i.e., ≤ 1%) can improve considerably the strength, durability, thermal and piezoresistive properties of AAM [57-61]. In particular, Ranjbar et al. [55] claimed that the compressive and flexural strength of fly ash geopolymer increased by 1.44 and 3.16 times respectively, corresponding to the inclusion of 1% GNPs. This was corroborated by Sajjad et al. [62-64], whose report indicated that the use of graphene dosage less than 1% correlated to the improvement in mechanical strengths of ambient-cured AAM by approx. 26 – 41%.

Although some preliminary studies as aforementioned displayed the promising mechanical outcome due to the incorporation of graphene derivatives in geopolymer matrices, investigations into the high-temperature stability of AAM incorporating carbon-based nanomaterials has been limited and remains unclear. Hence, this study aims to develop an ambient-cured nano-engineered AAM with excellent thermal characteristics. Different AAM mixes were synthesised from Class F fly ash (FA), ground granulated blast furnace slag (GGBFS), silica fume (SF), GNPs and sodium metasilicate (SMS) – a solid alkaline activator. After being subjected to 200ºC, 400ºC, 600ºC and 800ºC, AAM mixes were characterised for their thermal properties (i.e., thermal lag, specific heat) and residual compressive strength. Phase assemblage, chemical components and pore structure were also characterised via TGA/DTG, isothermal calorimetry, XRD, SEM/EDS and X-ray μCT to analyse the mechanisms behind the thermo-physio-mechanical stability of AAM. These provide a practical viewpoint on the enhanced AAM as a binding agent, developed for high-temperature TES technology.

## 2. Materials, mix designs and preparation

### 2.1 Materials

Class F FA, GGBFS and densified SF were the main precursors in this study. As determined through X-ray fluorescence (XRF), the primary constituents of FA are alumina and silica, meanwhile main components of GGBFS are calcium oxide and silicon dioxide, and those of SF is mostly silica. Their detailed chemical oxides/crystallinity is provided in Table 1. Particle size distribution and phase identity of these aluminosilicate precursors, as determined by Laser diffraction and X-ray diffraction (XRD) respectively, are presented in Fig. 1. The XRD results of the precursors indicated the amorphous nature of GGBFS and SF (> 70%) as well as major crystalline peaks in GGBFS (i.e., gypsum and calcite) and SF (i.e., quartz and moissanite). Meanwhile, multiple crystalline peaks of quartz, calcite, hematite, magnetite and mullite were detected in FA. Furthermore, the FA ($g_d$ = 2.4 g/cm$^3$) supplied by Cement Australia and the GGBFS ($g_d$ = 3 g/cm$^3$) contributed by the Independent Cement and Lime Pty Ltd had a relatively small median size $d_{50}$ = 22 mm and $d_{50}$ = 14 mm respectively. In contrast, the densified SF ($g_d$ = 2.27 g/cm$^3$) supplied by Microsilica Pty Ltd had a lager median size $d_{50}$ = 86 mm. The solid SMS purchased from Redox Ltd had a molar SiO$_2$/Na$_2$O of 0.9, density of 1.2 g/cm$^3$ and median particle size of approx. 1 mm. The GNPs in a powder form, supplied by First Graphene Ltd, possessed a maximum lateral size of 50 mm. The superplasticiser (SP) – Master Glenium SKY 8379, was used for improving the dispersion and workability of mixes.

Table 1. Mineralogical composition and crystallinity of aluminosilicate precursors

| Chemical composition (wt.%) | FA | GGBFS | SF |
|---|---|---|---|
| $SiO_2$ | 52.75 | 31.17 | 97.57 |
| $Al_2O_3$ | 27.66 | 11.86 | 0.69 |
| $Fe_2O_3$ | 8.12 | 0.3 | 0.07 |
| CaO | 6.66 | 44.34 | 0.19 |
| MgO | 1.23 | 5.57 | 0.74 |
| $K_2O$ | 0.85 | 0.41 | 0.51 |
| $TiO_2$ | 1.51 | 0.62 | – |
| MnO | 0.11 | 0.28 | 0.01 |
| $SO_3$ | 0.52 | 5.13 | 0.08 |
| $P_2O_5$ | 0.41 | – | 0.06 |
| LOI | 0.18 | 0.14 | 0.08 |
| Crystalline | 56.1 | 19.7 | 30 |
| Amorphous | 43.9 | 80.3 | 70 |

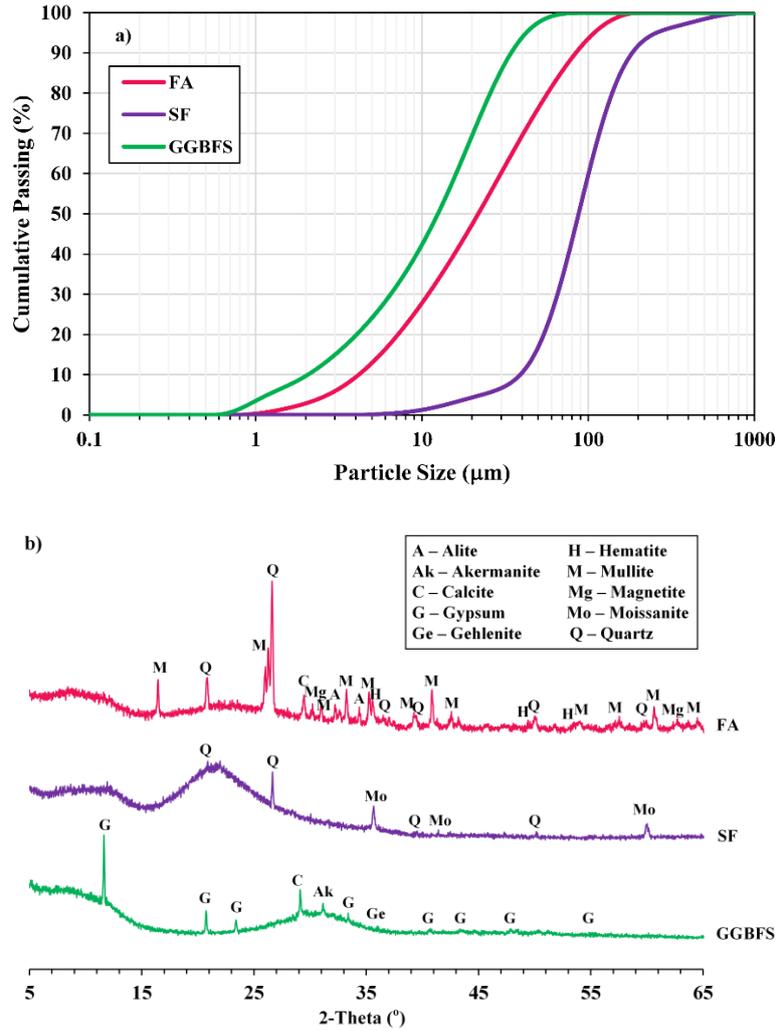

Fig. 1. Characteristics of aluminosilicate precursors: a) The particle size distribution, b) Phase identity using XRD

## 2.2 Mix designs and sample preparation

Five mixes with different unary, binary and ternary blends were designed to compare their performance at elevated temperature. For F and S samples, 100% FA and 100% GGBFS were used as the reference. FS sample was the binary combination of FA (60%) and GGBFS (40%). Meanwhile, FSSF and FSSFG samples comprised the ternary blend (55% FA + 40% GGBFS + 5% SF). Only FSSFG sample included further 0.1% GNPs. For more details, the mix proportions and molar ratios were displayed in Table 2. In all mixtures, the water-to-binder ratio was kept constant at 0.35. The mass ratio of solid activator to the precursors was fixed to 0.1 and the SP was used at 1% of the total weight of precursors.

Table 2. The mix proportions and molar ratios of AAMs

| Sample ID | Materials (kg/m$^3$) | | | | | | | Molar ratios | | |
|---|---|---|---|---|---|---|---|---|---|---|
| | FA | GGBFS | SF | GNPs | SMS | SP | H$_2$O | SiO$_2$/Al$_2$O$_3$ | Na$_2$O/Al$_2$O$_3$ | H$_2$O/Na$_2$O |
| F | 1164 | 0 | 0 | 0 | 116 | 12 | 407 | 2.1 | 0.2 | 7.0 |
| S | 0 | 1250 | 0 | 0 | 125 | 13 | 438 | 3.0 | 0.4 | 7.0 |
| FS | 718 | 479 | 0 | 0 | 120 | 12 | 418 | 2.3 | 0.2 | 7.0 |
| FSSF | 657 | 478 | 60 | 0 | 120 | 12 | 418 | 2.5 | 0.2 | 7.0 |
| FSSFG | 657 | 478 | 60 | 1.2 | 120 | 12 | 418 | 2.5 | 0.2 | 7.0 |

One-part alkali-activated pastes were prepared by pre-mixing dry constituents, including aluminosilicate precursors and solid SMS, at slow speed (61 Rpm) for 5 minutes. The aqueous solution of water and SP were prepared separately. For the mix incorporating nanomaterials, GNPs powders were ultrasonically treated in the SP/water solvent for 20 minutes using the sonicator Q500 to attain a good dispersion before slowly adding to the dry mix. To minimise the undissolved solid activator particle, the mixing duration was continued for further 15 minutes. In between, the high-speed mixing (113 Rpm) was operated for 1 minute at 4, 9 and 14 minutes after the water addition to ensure the uniform dispersion. Subsequently, the homogeneous slurry was casted into the 50 × 50 × 50 mm$^3$ cube moulds. These newly-casted mixtures were compacted with a support of vibrating table for 1 minute to eliminate entrapped air voids [65]. Plastic sheets were applied onto the surface of specimens to prevent shrinkage-induced cracks, as suggested [66]. After 24-hour storage under the laboratory condition (T = 24 ± 2ºC, RH = 50 ± 2%), those specimens were demoulded and kept inside the plastic bag until the test date. Notably, this ambient curing method had no signs of efflorescence on all AAM samples.

For SEM/EDS characterisation, a small section (10 × 10 × 10 mm$^3$) was cut out from broken parts of AAMs after mechanical testing. The SEM sample preparation procedures (i.e., grinding and polishing) was followed other studies [67-69]. For TGA and XRD analysis, the post-tested AAM piece was also retrieved and ground into fine particles (< 65 mm) with a support of Mill McCrone instrument. For specific heat test, fine AAM powders were also used. All samples prepared for microstructural characterisation, thermal and chemical analysis were pre-treated with acetone to eliminate the moisture content [66], thereby impeding the ongoing reaction.

## 3. Experimental methodology

The Tetlow furnace with a maximum temperature of 1300°C was used in this investigation. Except for the specimen cured at ambient temperature as the reference, other AAM samples were subjected to elevated temperatures at 200°C, 400°C, 600°C and 800°C with a heating rate of 10°C/min. Those temperature was selected, corresponding to each stage of thermal degradation of AAMs, including moisture evaporation, phase decomposition, calcination and sintering. When the heat reached the targeted temperature, it was maintained for 2 hours to ensure the whole sample was heated up evenly. Afterwards, the AAM samples were naturally cooled down to room temperature. During the cooling period, the thermal digital camera was utilised to track and record the reducing temperature of AAM specimens. Those data were used to estimate the thermal lag of AAM specimens. The heating regime and temperature tracking procedures are illustrated in Fig. 2 below. For visual representation, a certain surface of sample from each group was captured by camera to detect cracking after exposing to different temperatures.

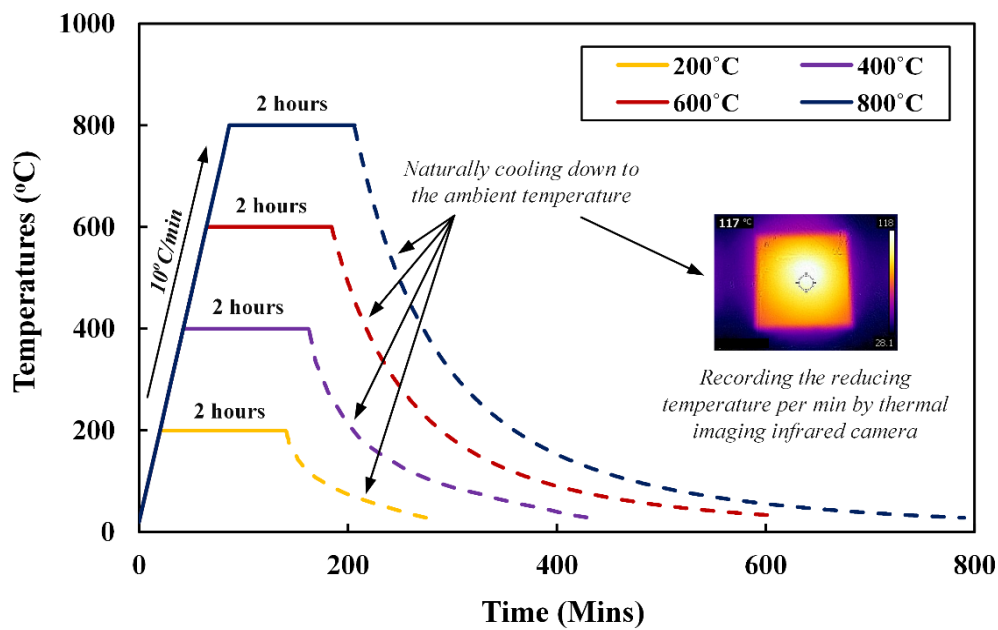

Fig. 2. Heating regime and temperature tracking of AAM samples at elevated temperatures

For each mechanical data point, four standard specimens were tested, and the mean value was reported. The compressive strength of samples was undertaken at 28 days by Technotest machine with loading rate of 1kN/s, according to ASTM C109/C109M-20b. The thermogravimetric analysis (TGA) was conducted using a PerkinElmer TGA 8000 in the $N_2$ atmosphere with a purge rate of 20 mL/min. The temperature range of TGA was programmed between 35ºC and 850ºC with a heating rate of 10ºC/min. SEM/EDS analysis was conducted through the FEI Teneo VolumeScope, operating at a working distance of 10 mm and a voltage of 10 kV. XRD patterns were obtained using a Bruker D8 Advance Powder Diffractometer. The grounded powder samples were passed through a 65 mm-mesh sieve before placing into the sample holder with flattened surface. Diffraction analyses were set up to run from 5º to 65º at 2θ using copper K radiation. The excitation voltage and current were 40 kV and 40 mA with a counting time of 1 sec and wavelength of 1.5418 Å.

The hydration kinetics of AAM paste were assessed by using a TAM Air 8-channel isothermal calorimeter. About 40 g of each AAM paste was manually mixed and placed into an ampoule with a sealed cap before loading into a channel of calorimeter. The temperature of the calorimeter was kept at 23 °C during 24 hours of testing. The recorded heat flow and cumulative heat of hydration over time were normalised with respect to the total mass of each AAM sample for the comparative purposes. The Hot Disk TPS2500, which was connected to a planar Kapton sensor embedded to a Cp cell, was used to record and measure the specific heat of AAM specimens in the powder form.

Micro-CT (μCT) was collected using a Phoenix Nanotom M (Waygate Technologies) operated using xs control and phoenix datos|x acquisition software. Full samples of casted cube (50 × 50 × 50 mm$^3$) were mounted in the instrument and 15-min scans were conducted collecting 1798 projections through a full 360-degree rotation of the specimens at a resolution of 42.5 micrometer. An x-ray energy of 140 kV and 300 microAmp was used with a 0.25 mm Cu filter to preharden the x-ray beam. Volume reconstruction of the micro-CT data was performed using the phoenix datos|x reconstruction software applying an inline median filter, ROI filter and beam hardening correction filter. Volume reconstructions were imported to Avizo (Thermo Fisher Scientific) for analysis. A combination of interactive thresholding and interactive top-hat thresholding was used to segment the internal porosity and cracks forming within the concrete cubes. A volume fraction tool was used to determine the % porosity distribution in cross-sectional slices through samples. A pore size distribution was determined by running a label analysis on the segmented pore structure to determine the volume and equivalent spherical diameter of separate pores.

## 4. Results and discussion

### 4.1 Visual observation

The physical appearance of AAM specimens after high temperature exposure was monitored to observe the colour change and crack behaviour (Fig. 3). The visible crack and discoloration of the sample surface was pronounced with increasing temperatures, especially at 800°C where all mix designs exhibited multi-cracks, thermal deformation and discolouration. Severe cracking was detected in S specimens, in which a sign of fine cracks started appearing at 200°C and further turned into macro-cracks and fragmentation after being subjected to 400 – 800°C. The teal colour of S specimens, caused by media containing sulphur [70], was gradually faded into light grey at up to 800°C. For F samples, few micro-cracks were initiated at 400°C, and abruptly propagated into multi-branched macro-cracks when reaching to 800°C. The colour on their surface also turned to reddish at this high temperature, which could be attributed to the oxidation of iron compound presented in FA ($Fe_2O_3 > 8\%$). Noticeably, FS and FSSF specimens achieved a relatively good thermal stability up to 600°C. At 800°C, FS and FSSF samples displayed macro-cracks and physical deformation due to high thermal shrinkage, but they showed less severe cracks than S samples. When incorporating GNPs, FSSFG specimens displayed a grey colour and had few microcracks on the surface from 400°C, which was then slightly branched at 600°C and further expanded when reaching to 800°C. It could be realised that the FSSFG specimens with GNPs were prone to have more cracks at high temperatures, compared to the FSSF. Notably, upon attainment of a calcination temperature of 800°C, the presence of micro-fissures signified the decomposition of the amorphous phase at elevated temperatures, transitioning towards a crystalline phase, indicative of the potential commencement of sintering [71].

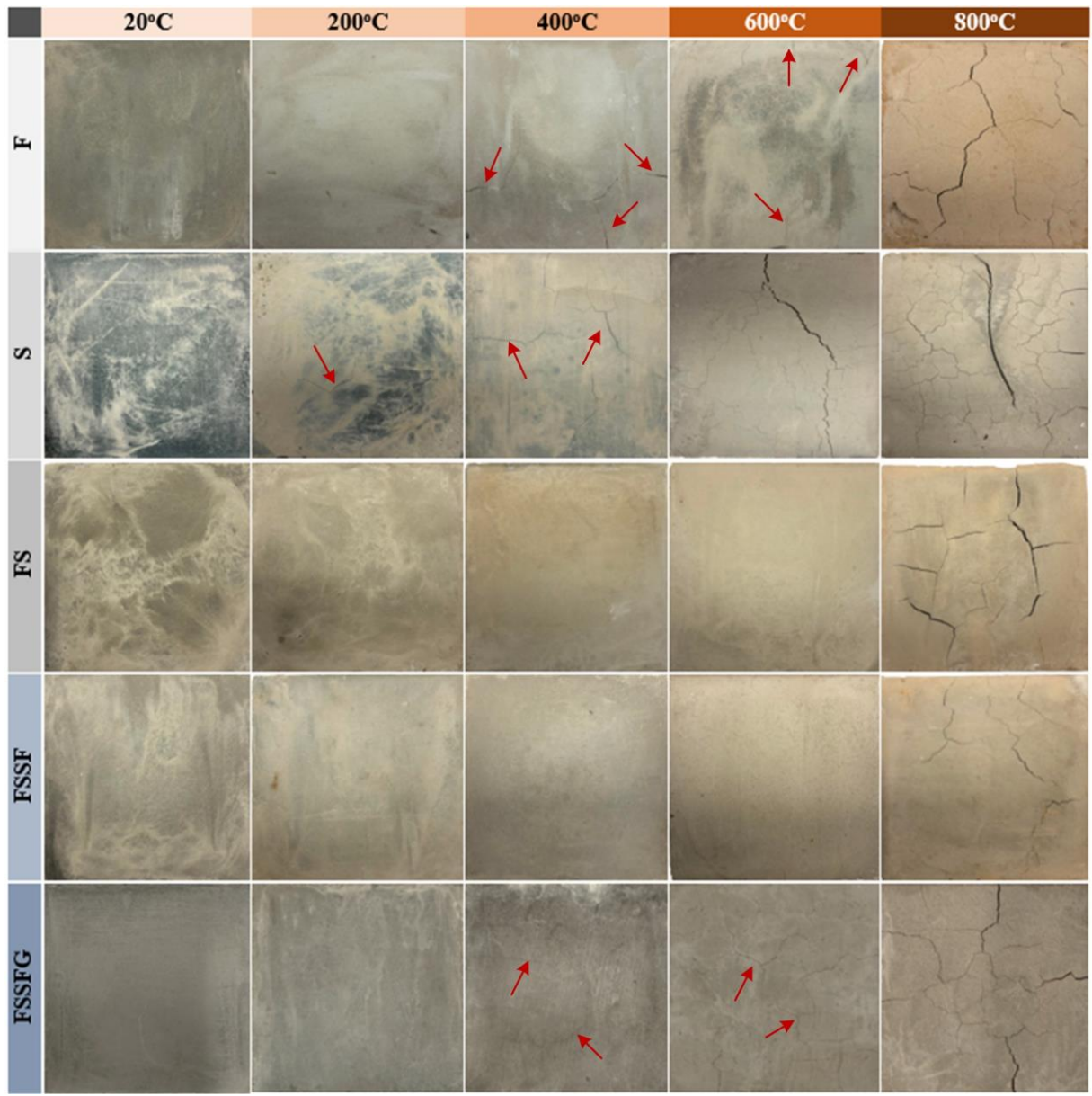

Fig. 3. Visuals of different AAM samples before and after exposing to elevated temperatures. Red arrows indicate microcracks on the surface of samples.

## 4.2 Residual strength

The variation in compressive strength of AAM pastes at high temperatures with trendlines and equations are shown in Fig. 4. It was observed that F samples achieved the lowest 28-day strength (< 10 MPa). Lower Si/Al ratio of F samples compared to that of others, together with immature aluminosilicate bonds at ambient condition could be the reason for their lowest strength (see Fig. 12). As reported in other studies [72, 73], geopolymers with lower Si/Al ratio often exhibited lower strength despite a better thermal stability. For those AAM mixes with slag, calcium-rich content drove the formation of C-A-S-H at the expense of N-A-S-H. Thus, their post-heated strengths were increased due to the coexistence of C-A-S-H and N-A-S-H gels as well as the increase of Si-O-Si bonds (higher Si/Al ratio) in the N-A-S-H gel [72]. Apparently, S specimens reached to the highest 28-day compressive strength with over 80 MPa under ambient condition. With the incorporation of GNPs, FSSFG samples could obtain an equivalent strength to S specimens, meanwhile FSSF and FS samples exhibited a negligible difference in strength (~ 74 MPa) at 28 days.

Interestingly, high heat and build-up vapour pressure at elevated temperatures were likely to establish the autoclaving condition inside AAM pastes, which facilitated further geopolymerisation process for strength improvement. Also, the expelled water from the geopolymer matrix during high temperature exposure could result in a discontinuous nano-pore network, thereby enhancing the strength of geopolymers [74]. In contrast, thermal stress and build-up vapour pressure somewhat diminished this positive effect and triggered strength loss [75]. Those contradictory hypotheses clarify the variation in compressive strength of AAM pastes at high temperatures. It should be noted that F samples were recorded as having a 15 – 105% strength gain after exposing from 200°C up to 800°C despite multiple cracking observed (see Fig. 3). A similar trend for the increased strength of geopolymer was found in another report, which was somewhat ascribed to the solidification of melted phases during the cooling process [76]. S specimens also gained strength of approx. 100 MPa (~ 20% increment) upon exposure to 200°C. Nonetheless, they were vulnerable to higher temperatures of 400 – 800°C that led to a huge strength loss by approx. 47 – 87% and remained the lowest strength (around 12 MPa) among investigated AAM mixes at 800°C. Such considerably low residual strength of S samples at elevated temperatures was mainly attributed to the brittle nature of C-(A)-S-H susceptible to thermal deformation, dehydration and crystallisation [77, 78]. Meanwhile, FS, FSSF and FSSFG specimens were more stable up to 400°C with an increasing compressive strength recorded. Especially, FSSFG samples obtained 28-day compressive strength of 103.7 MPa (28.6% improvement) at 400°C, followed by the FSSF with similar strength of 100.8 MPa (32.1 % improvement), those achieved the highest strength gain compared to other AAM mixes. It should be highlighted that the excellent thermo-mechanical performance of FSSF and FSSFG mixes was achieved at 400°C, attributed to the additional formation of highly cross-linked C-(A)-S-H binding gel to densify the microstructure [78]. As temperature rose to 600°C, the decomposition of phases and thermal cracking enlargement (see Fig. 8 and Fig. 13) triggered the strength degradation of FS, FSSF and FSSFG samples, especially a noticeable reduction in strength by 54 – 65% was recorded when being subjected up to 800°C. Only FSSFG samples could retain the highest residual strength of 37.4 MPa at 800°C.

The upward trend in compressive strength of AAM pastes over elevated temperature exposure was in agreement with other published studies [79-82]. Such heat effect on strength development after high temperature treatments could be dependent on the retained water in microstructure and the availability of alkaline sources [81]. This phenomenon also indicated that the strength of ambient-cured AAM pastes had not been fully developed yet until being activated for further geopolymerisation and hydration under heat treatment. This could be the reason why AAM mixes with pre-heat curing were reported as having higher compressive strength at ambient condition but lower strength after exposing to elevated temperatures (> 200°C). Findings in the study of Zhang et al. [83] also corroborated this observation, as reported that the ambient-cured geopolymer achieved better strength enhancement and lesser strength deterioration at elevated temperatures than the heat-treated ones.

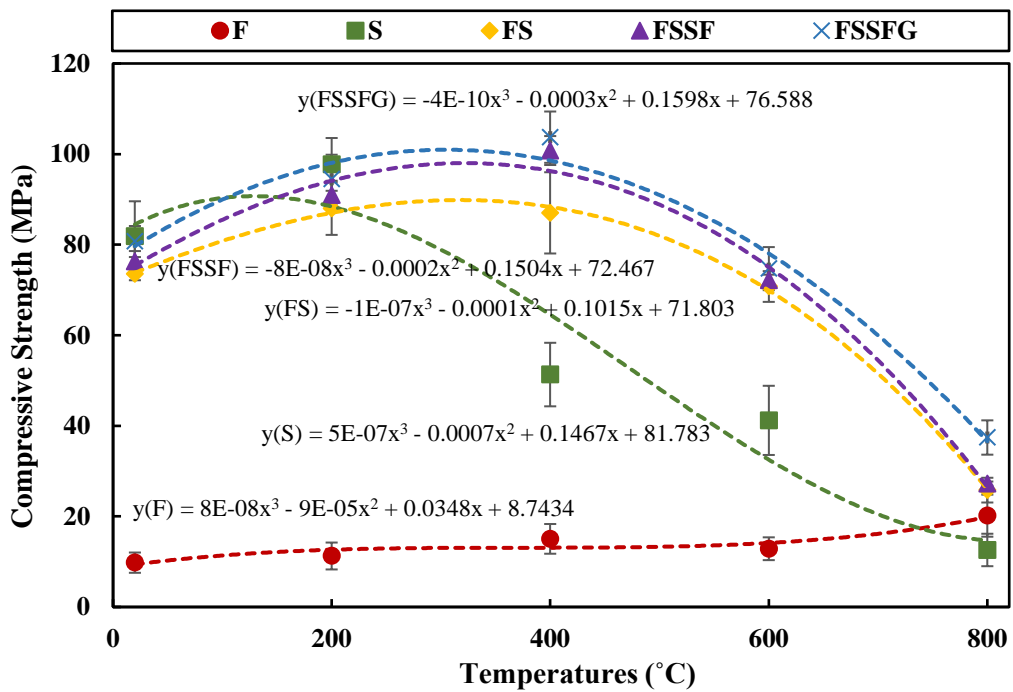

Fig. 4. Compressive strength of AAMs before and after exposing to elevated temperatures

### 4.3 Thermal lag

Thermal lag refers to the duration that the heat energy stored in material is released to the surrounding. This delay time in heat release is dependent on many influential factors such as the heat capacity, thermal conductivity, density and size of materials. Fig. 5 presents the thermal lag of AAM mixes at different temperatures. Overall, thermal lag of investigated AAM mixes with the thickness of 50 mm varied within 2.5 – 4.5 hours after temperature exposures in a range of 200 – 800ºC. It is exhibited that increasing temperatures correlated to higher thermal lag in AAM pastes. However, the increase rate in thermal lag at higher temperatures (600 – 800ºC) tended to be slightly less pronounced. The appearance of multi-cracks on their surface and shrinkage-induced size reduction could render heat transfer faster from the material's surface to the ambient and thus somewhat shorten the thermal lag. Among all, F samples obtained the shortest thermal lag; meanwhile, S specimens displayed the longest delay time to maintain the heat. This result was reasonable as S samples possessed a larger amount of calcium carbonate to produce latent heat than other AAM mixtures (see Fig. 8 and Fig. 9). Among all, the thermal lag of FS, FSSF and FSSFG samples seems to be equivalent.

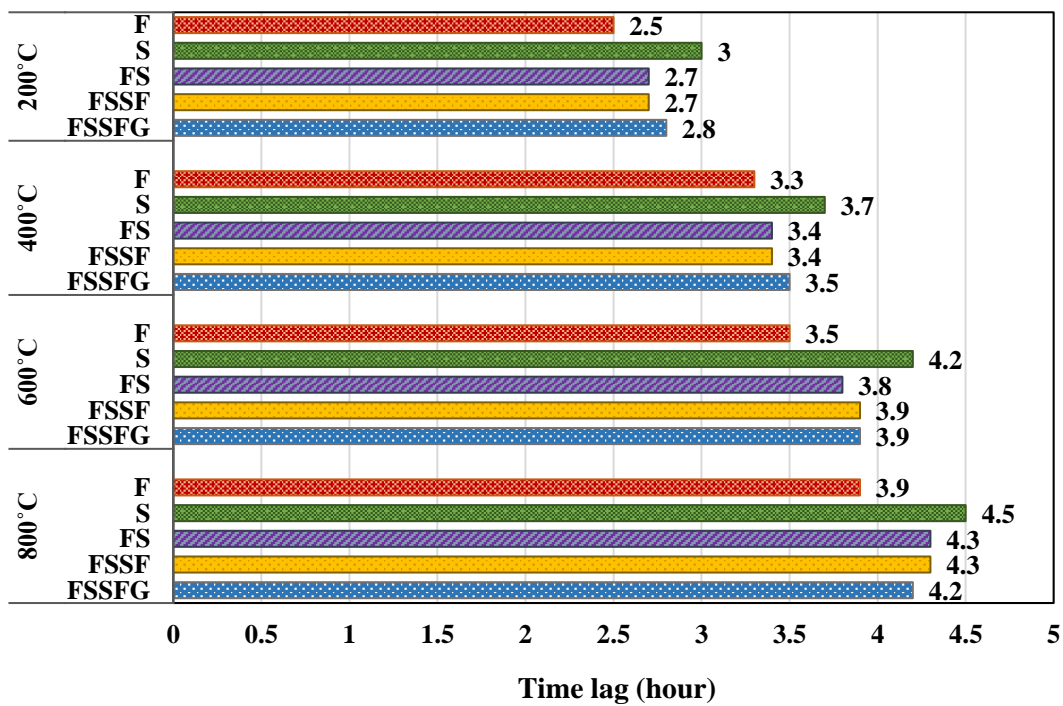

Fig. 5. Thermal lag of AAM samples corresponding to different high temperature exposure.

## 4.4 Specific heat capacity

Interestingly, there was a downward trend in the specific heat capacity ($C_p$) of AAM when exposing to elevated temperatures, as displayed in Fig. 6. This was opposite to the observation of Jacob et al. [84], who used differential scanning calorimeter (DSC) method to measure the $C_p$ value and observed an increasing tendency of $C_p$ value when exposing to high temperature.

It should be noticed that the DSC test in the study of Jacob et al. [84] computed the $C_p$ value at specific temperatures from the heat flow-temperature curve under constantly increasing temperatures. On the other hand, the $C_p$ values in this study was measured from residual samples in powder forms after exposing to 200 – 800°C. Specific heat capacity of AAM samples reduced after elevated temperature exposure could be ascribed to water removal and increased crystallinity of phases (see Fig. 8 and Fig. 10). Among all, F samples had the lowest $C_p$ value at around 700 – 900 J/Kg°C corresponding to 20 – 800°C exposure. F samples also experienced the least variation in $C_p$ value range, meanwhile other sample groups showed a significant reduction (over 40%) of $C_p$ value upon high temperature exposure. With the presence of slag, the heat capacity of AAMs was improved significantly, which was also corroborated by another study [85]. Notably, FSSFG samples exhibited the highest $C_p$ value over others, with up to ~ 1400 J/Kg°C at ambient condition. The $C_p$ values of S, FS and FSSF samples were almost equivalent throughout high temperature exposure. When exposing to 800°C, the $C_p$ value of all AAM were similar and below 800 J/Kg°C. From other results in the literature, the $C_p$ value of AAM paste at ambient temperature in this study was significantly higher than those of cement-based materials (700 – 820 J/Kg°C) [86-88] and heat-cured geopolymer paste (~ 850 J/Kg°C) [34]. As compared with other pioneer concrete developed for TES modules at 400°C, $C_p$ value of FSSFG paste ($C_p^{400}$ = 1120 J/Kg°C) in this study was lower than those of Heatcrete® ($C_p^{400}$ = 1280 J/Kg°C) [36], but higher than and DLR® concrete ($C_p^{400}$ = 1050 J/Kg°C) [89] and GEO concrete ($C_p^{400}$ ~ 950 J/Kg°C) [90]. Notably, the developed AAM paste still achieved an equivalent-to-higher $C_p$ value than concrete reported in previous studies despite a low density without aggregate. This result indicated the promising use of AAM paste from this study for TES systems with the operation temperatures at around 400°C.

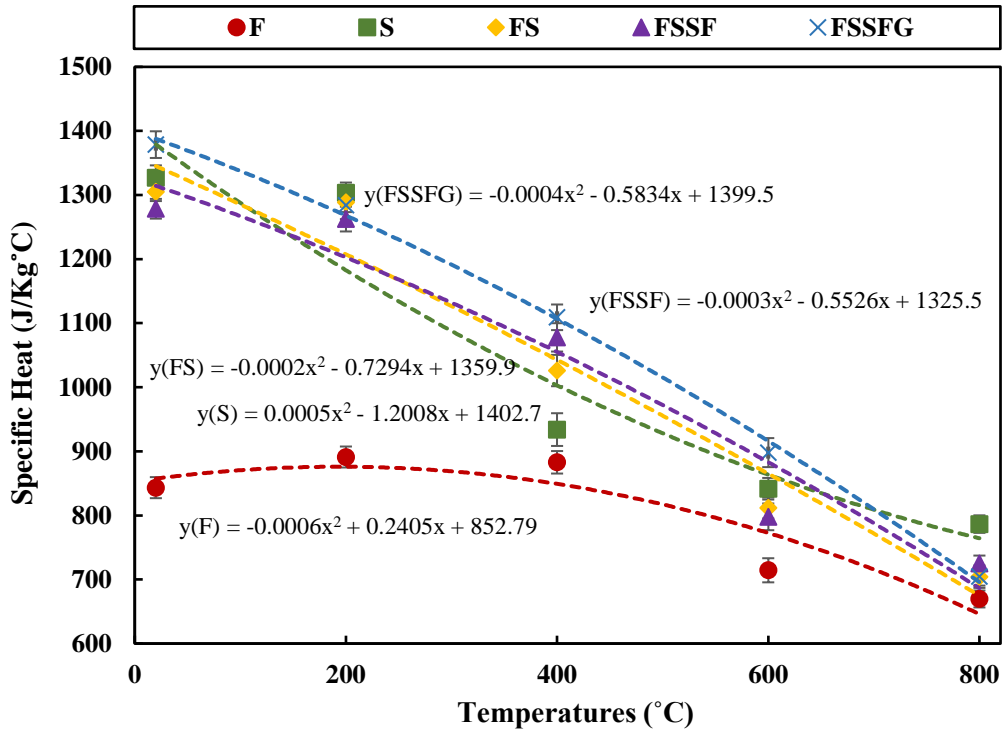

Fig. 6. Specific heat ($C_p$) of AAM specimens at elevated temperatures

## 4.5 Isothermal calorimetry responses

Fig. 7 presented the heat evolution over 24 hours for five AAM mixtures. During the first hour, a sharp peak was observed for all AAM samples, indicating the heat released due to the dissolution of particles during the induction period. This first exothermic peak was significant with the addition of GNPs (FSSFG samples), followed by the S sample. Whereas, F samples exhibited a relatively low peak during the first dissolution stage. A low volume fraction of amorphous phase of FA resulted in a less vitreous structure and less dissolved species for the subsequent reaction process. This phenomenon explicitly correlated to the lowest strength of F samples at ambient condition (see Fig. 4). The second peak reflected the main reaction process, associated with the precipitation and polycondensation of silicate and aluminate species. Noticeably, there was hardly any sign of second peak observed from F samples. In contrast, high $Ca^{2+}$ contents in S specimens were prone to boost the main reaction peak in the acceleration periodt, which was also reported in another study [91]. This can be assigned to the breakdown of the covalent Si-O-Si and Si-O-Al bonds in the amorphous phase of slag to release of these ions into a highly alkaline pore solution, which facilitated the formation of C-(A)-S-H gels [92]. Also, the presence of GNPs provided pseudo-contact points for forming N-A-S-H and C-(A)-S-H bridges, which could result in a further heat release in this stage.

After the second peak, there was no other noteworthy peak as the heat of reaction was prone to gradually decelerate. This was due to the growth of C-(A)-S-H gel layer on the surface of precursors to form the reaction product shell, by which the $OH^-$ ions crossing the shell was hindered at a very low speed and then the reaction gradually stopped [93]. During 24-hour reaction, the S samples recorded the highest cumulative heat release (107.4 J/g), followed by the FSSFG (98.4 J/g), FSSF (95.2 J/g) and FS samples (92.6 J/g). The lowest heat release was observed from the F sample with only 25.9 J/g. It was noticed that the increasing of silica content (i.e., addition of SF) slightly broadened the peak of heat release and cumulative heat generation of mixtures, as observed from the FSSF and FSSFG in comparison to FS samples.

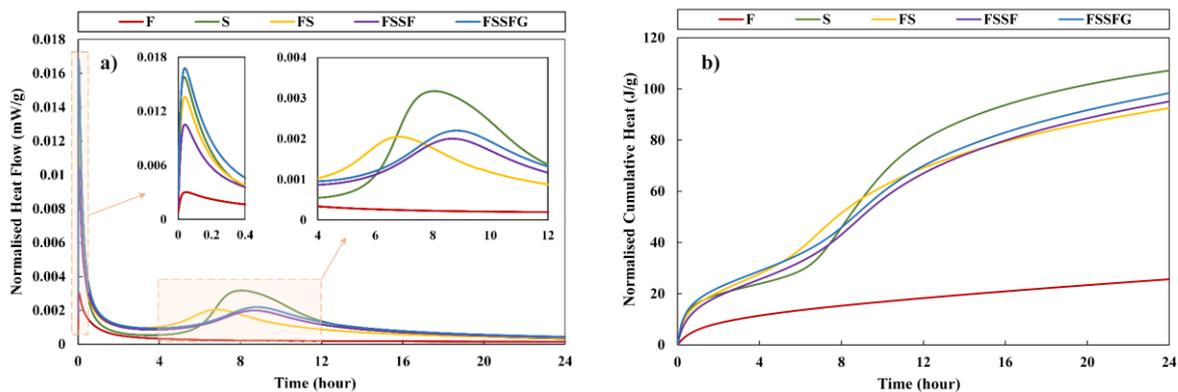

Fig. 7. Reaction heat flow (a) and cumulative heat (b) of different AAMs within 24 hours

## 4.6 TG/DTG analysis

The relative gel content and thermal stability of ambient-cured one-part AAM pastes can be reflected by TGA/DTG curves, as displayed in Fig. 8. Commonly, there are three main stages where the mass change takes place: i) the first stage up to 230°C (> 60% mass loss) involves the major evaporation of capillary water and physically bound water from N-A-S-H and C-(A)-S-H gels; ii) the second stage within 230 – 700°C (mild change) is assigned to the volatilisation of chemically-bonded water, loss of $CO_2$ from vaterite and dehydroxylation of hydroxyl groups; iii) the third stage above 700°C (almost remained unchanged) is attributed to the decomposition of calcite, crystalline phase transition from C-(A)-S-H gels to other phases [83]. From the TGA/DTG curve, the mass loss initiated in the very early stage (30 – 40°C) due to the dehydration of Natron ($Na_2CO_3 \cdot 10H_2O$) into sodium carbonate heptahydrate and monohydrate [94]. The significant mass loss occurred within 60 – 75°C could be assigned to the evaporation of free water and continued dehydration of Natron. Major reduction in mass within this temperature range was observed from S, FS and FSSF samples, indicating that the carbonation was more pronounced in these AAM mixes with high Ca content compared to F samples. Interestingly, FSSFG mixes with GNPs showed the lowest mass loss, which implied the effect of GNPs on the formation of less carbonation products.

In the temperature range of 100 – 230°C, the physically-bound water was constantly evaporated for condensation process followed by the formation of geopolymeric binding gels, thereby further amplifying mass loss. Such weight loss in the first stage could somewhat reflect the degree of geopolymerisation since it was closely associated with the formation of N-A-S-H and C-(A)-S-H gels. In addition, the shallow peak appeared around 350°C for S, FS and FSSF samples, which was ascribed to the liberation of interlayer water of geopolymeric gels during geopolymerisation. This phenomenon also correlated to the high strength of FS and FSSF mixes at this temperature range (see Fig. 4), except for S samples due to crack appearance (see Fig. 3). The dehydroxylation of $OH^-$ groups also took place within the temperature range of 250 – 700°C for different mixes and thus caused another minor peak of mass loss during the second stage. Above 700°C, the mass loss rate of samples was deemed to be stabilised. CaO-rich S samples still experienced a gradual reduction in weight with the increase of temperatures. The peak at around 750 °C was associated with the decarbonation of calcite and rendered S samples less thermally stable at elevated temperature than the other AAMs. Also, the appearance of nepheline phases (see Fig. 9) somewhat contributed to a slight mass loss of F matrix at 750 °C [95]. When heating up to 800°C, not only the loss of binding water in C-A-S-H gels but also the re-formation of newly-ordered crystalline phases from the destroyed gel structure would take place [96].

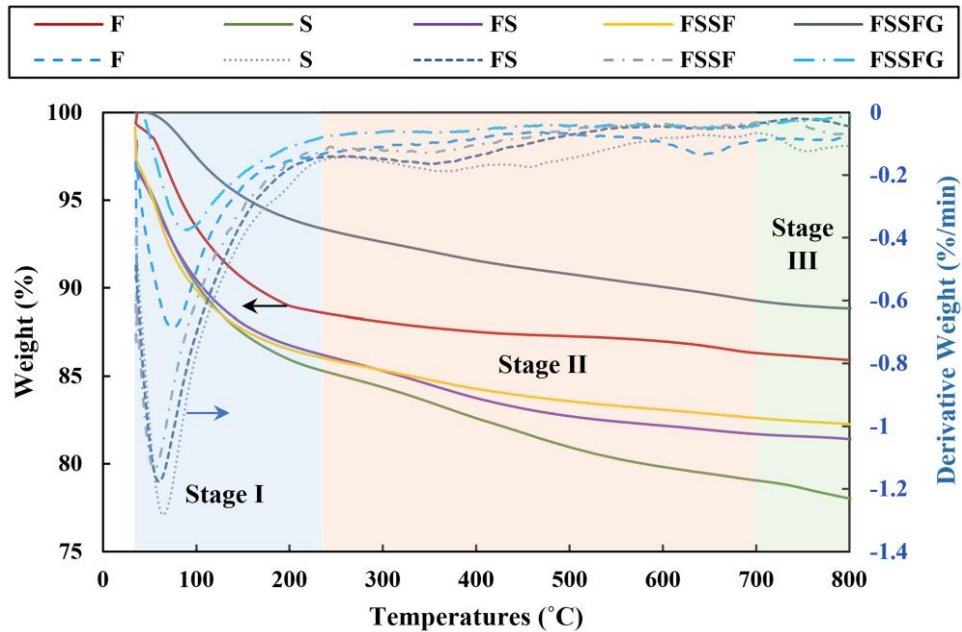

Fig. 8. TG/DTG of the ambient-cured one-part AAM pastes. The solid lines denote Weight (%) and the dash lines denote Derivative Weight (%/min)

## 4.7 X-ray Diffraction (XRD)

The XRD diffractograms of one-part AAM pastes at high temperatures are presented in Fig. 9. For F samples, a diffuse band positioned between 20–35° 2θ, especially the sharp peak of quartz at 26.6° 2θ, was associated with the development of alkali-silicate glasses and the presence of amorphous aluminosilicate gel network. The main crystalline phases originally contained in the raw FA, including quartz ($SiO_2$), mullite ($3Al_2O_3 \cdot 2SiO_2$), magnetite ($Fe_3O_4$) had little change after exposure up to 600°C. This indicated that these phases had relatively high resistance to elevated temperatures. When F samples reached to the sintering temperature (800°C), the onset of high-temperature oxidation of $Fe^{2+}$ ions proceeded the transformation of magnetite into hematite ($Fe_2O_3$). At 800°C, there was also an emergence of new crystalline phases, namely diopside ($CaO \cdot MgO \cdot 2SiO_2$), hedenbergite ($CaO \cdot FeO \cdot 2SiO_2$), aegirine ($[Na,Fe]Si_2O_6$) and nepheline ($[Na,K]AlSiO_4$). The formation of the diopside and nepheline was the result of the phase transformation from silica-rich gels and alumina-rich gels respectively during the sintering process [97]. Regarding S sample, the reaction product was presented near 29.5° 2θ, resembling the diffraction pattern of a disordered C-S-H type gel with a riversideite structure, in which $Si^{2+}$ was partially substituted with $Al^{3+}$ referred as C-(A)-S-H – the main phase in alkali-activated slag. The intensity of amorphous hump at approx. 25–35° 2θ, along with traces of calcite, was gradually diminished by exposure up to 600°C.

Nonetheless, it was no longer identifiable upon exposure at 800°C, indicating that amorphous C-(A)-S-H gels had fully dehydrated or crystallised. This phenomenon corroborated the observation from TGA results (see Fig. 8) and correlated to the sudden drop in residual strength of S sample after exposure to the sintering temperature (see Fig. 4). Interestingly, although the crystalline phases of akermanite and gehlenite, contained in the raw GGBFS (see Fig. 1b), disappeared during the geopolymerisation upon 600°C, they started to reappear in the XRD pattern at 800°C. Those phases were frequently associated with the crystallization of Ca-rich AAMs treated at elevated temperatures [98]. The structural instability of aluminosilicate gels at elevated temperature resulted in the formation of crystalline gehlenite ($Ca_2Al[AlSiO_7]$), akermanite ($Ca_2Mg[Si_2O_7]$), zoisite ($Ca_2Al_3[SiO_4][Si_2O_7]O(OH)$) and merwinite ($Ca_3Mg[SiO_4]_2$), along with an appearance of crystalline nepheline at 800°C, which was also reported by Park et al. [82]. As observed, high Ca/Si ratio of the S samples (see Fig. 12) could lead to the dominant formation of gehlenite after exposure to 800°C [78]. Also, it should be noted that the presence of crystalline akermanite could be associated with highly porous microstructure, as observed from SEM and X-ray μCT (see Fig. 11 and Fig. 13) as well as reported in other studies [78, 99].

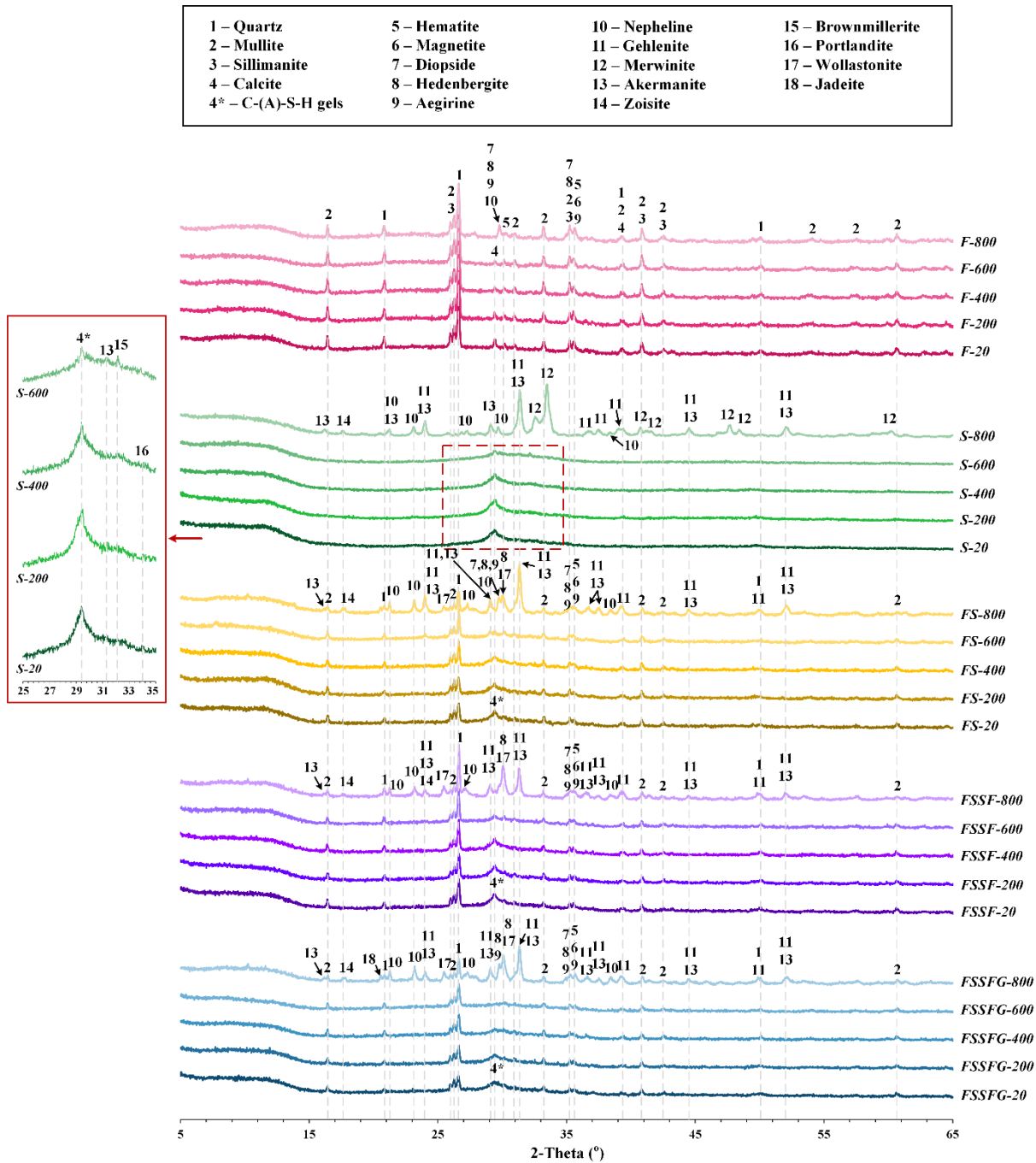

Fig. 9. XRD patterns and phase identity in different AAM mixes at elevated temperatures

When combining FA and GGBFS, the intensity of C-(A)-S-H peak in FS samples was much lower than that of S samples. Due to lower slag content, reduced $Ca^{2+}$ availability in the FS matrix delayed the evolution of the C-(A)-S-H phase, indicating the alteration of their reaction kinetics. Indeed, the strength of FS samples was lower than that of S samples at ambient temperature (see Fig. 4). After exposure to 200°C and 400°C, the XRD patterns of FS paste remained almost unchanged and showed no obvious deterioration in this condition before reaching to 600ºC where the peak intensity of C-(A)-S-H was evidently reduced. A similar tendency was also reported in the study of Pan et al. [100]. It should be noted that the heat resistance of C-(A)-S-H was weaker than that of sulphoaluminate gels (e.g., quartz and mullite, etc...), thus the FS samples explicitly achieved higher thermal stability than CaO-rich S samples. At 800ºC, the XRD pattern shifted remarkably with multiple peaks, where they indicated the crystallisation of C-(A)-S-H gels into akermanite and gehlenite, as well as the crystallisation of N-A-S-H into nepheline [82, 101]. Notably, the characteristic peak of merwinite at 32.6º, 33.4º, 47.7º and 48.5º 2θ observed from S sample, did no longer exist in FS matrix at the same sintering temperature. Instead, an appearance of new peaks at 26º and 30º 2θ indicated the existence of wollastonite phase ($CaSiO_3$) due to the reaction between quartz ($SiO_2$) and lime (CaO) at sintering temperatures. As reported, the higher the quartz content, the lower the formation temperature of wollastonite, especially in the gehlenite-bearing system [102]. Despite the small amount of Si-rich SF (5%) used in FSSF matrices, their XRD pattern was fairly similar to that of the FS sample, except for a higher quartz peak intensity at 26.6º 2θ. Due to high Si content, high Si/Al ratio in the mixture (see. Fig. 12a) was the premise for the formation of wollastonite phase upon exposure to elevated temperatures, resulting in the amplified intensity peak at 30º 2θ at 800ºC. The XRD patterns of FSSFG samples (with GNPs) were not much different from those of FSSF samples, except for the rise of jadeite phase ($NaAlSi_2O_6$) at 800ºC.

Despite the similarity in phase formation of FS, FSSF and FSSFG samples, the crystallinity of each phase in their matrix was distinct at different elevated temperatures, causing the variation in the amount of amorphous and crystalline phase. As shown in Fig. 10, increasing temperature exposure accelerated the phase crystallisation. A similar trend was also reported in the study of Ponomar et al. [103], where the amorphous fraction declined sharply at extremely high temperatures due to the formation of akermanite and gehlenite. A high amount of amorphous phase was found in F samples; meanwhile, other mixes containing GGBFS displayed a higher fraction of crystalline phases. Noticeably, greater phase crystallisation was observed from FSSFG samples with GNPs, compared to FS and FSSF samples. This could be attributed to the nucleation effect of GNPs that provides the additional nucleating site for facilitating the geopolymerisation/hydration reaction and crystallisation of amorphous phases with further ordered structure. Presumably, such mechanisms triggered by sintering effect and densification at high temperatures strengthened FSSFG matrices so that they could achieve higher residual strength than other investigated AAMs over different elevated temperatures (see Fig. 4).

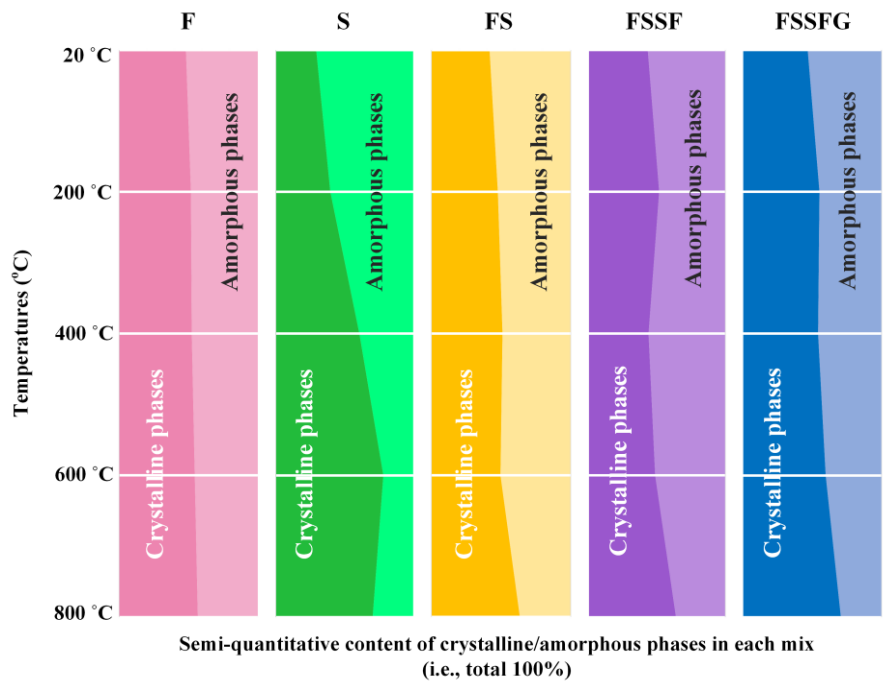

Fig. 10. Crystallinity in different AAMs at elevated temperatures

### 4.8 SEM/EDS

Microstructure of AAM samples characterised by backscattered electron imaging is displayed in Fig. 11. As observed, a large quantity of unreacted particles was spotted in the AAM mixes, especially at ambient temperature. It was justifiable that the geopolymer reaction of precursors was promoted by heat, thus a denser microstructure with less unreacted particles was recorded for AAM samples exposing to higher temperatures within 200 – 400°C. When exposing up to 800°C, porous matrix and crack defects induced by built-up vapour pressure were detected to be extensively growing. Noticeably, microcracks appeared in S specimens irrespective of temperatures, which turned into macrocracks and severe fractures from 400°C to 800°C. This could be due to the thermodynamic instability and low creep modulus of C-A-S-H phases - mainly formed in S samples, which triggered a significant shrinkage-induced cracks [104]. Although F specimens exhibited good thermal stability, a plethora of unreacted FA was widely distributed throughout the matrix and thus correlated to a relatively low residual strength (see Fig. 4). After an exposure up to 400°C, the microstructure of FS, FSSF and FSSFG samples were densified compared to those at ambient temperatures, which was also an implication for the strength improvement observed in Fig. 4. But more defects and porosity were found in their microstructure associated with strength reduction when the temperature reached to 600°C and 800°C. Notably, the inclusion of GNPs somewhat rendered its dense matrix susceptible to built-up vapour pressure at elevated temperatures. In specific, FSSFG specimens displayed more cracks than the FSSF despite a higher residual strength recorded.

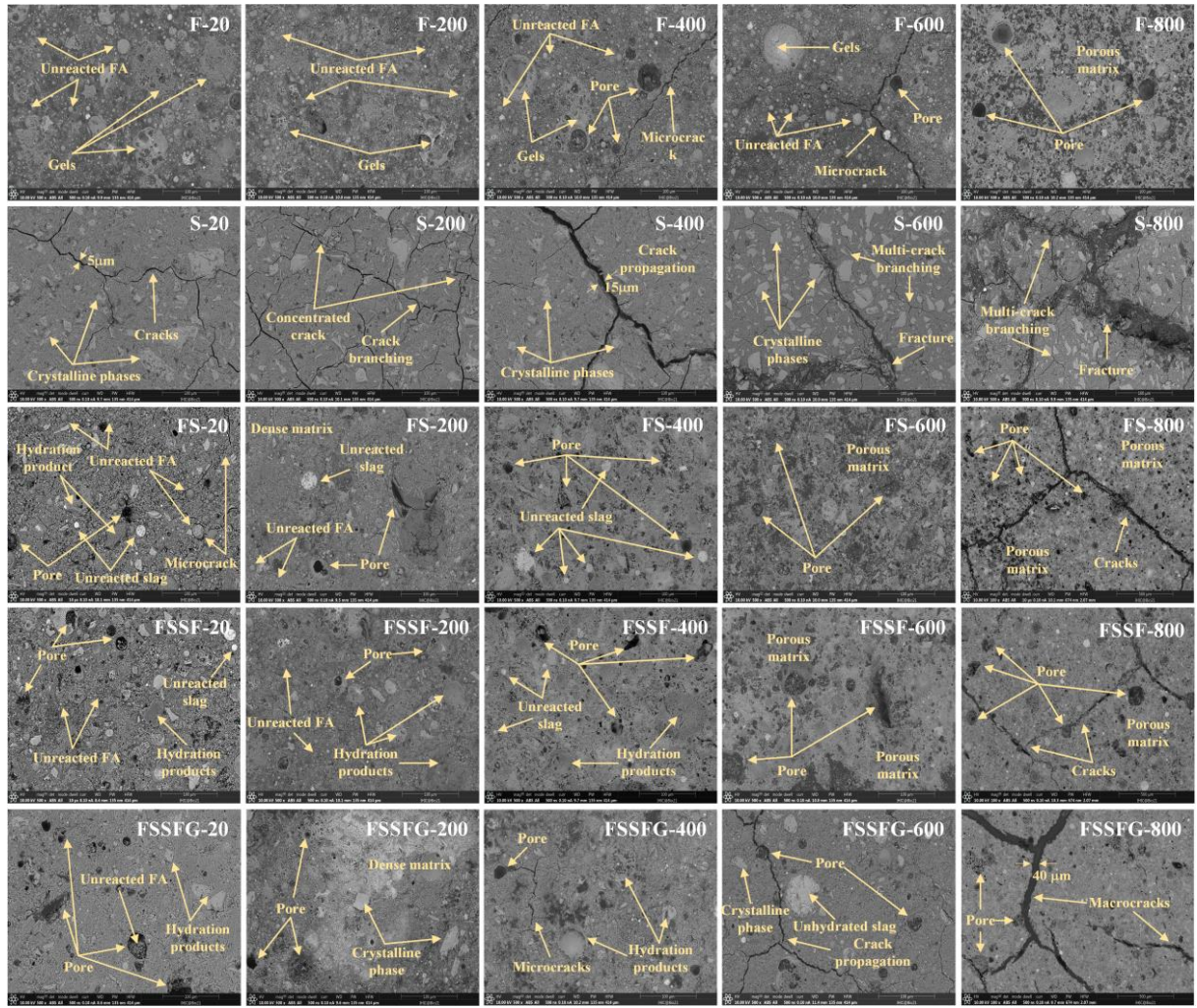

Fig. 11. Backscatter SEM micrographs of AAM mixes at various temperatures.

The elemental composition of gel products in AAM mixtures at elevated temperatures was detected and exhibited in two ternary Ca-Al-Si and Na-Al-Si plots together with their average Si/Al, Ca/Si, Na/Al and Na/Si ratios (Fig. 12). Due to a slight difference in mix design, those elemental ratios in the FS, FSSF and FSSFG samples varied within a similar cluster. The incorporation of $SiO_2$-rich SF contributed to a higher Si/Al ratio (> 3.0) in gels of FSSF and FSSFG specimens in comparison to FS samples. Interestingly, the Si/Al ratios in gel products of AAMs except for S sample tended to rise at the temperature up to 400$^o$C, indicating more C-(A)-S-H and N-A-S-H gels produced by heat-promoted geopolymerisation. This correlated to the increase in strength of ambient-cured AAMs at high temperatures in this study (see Fig.

4). However, non-reacted residual silica was prone to cause thermal instability (e.g., swelling) at elevated temperatures [105], which led to the severe cracking observed in Si-rich matrix with high Si/Al ratio (see Fig. 3 and Fig. 11). Despite a low Si/Al ratio recorded in gel systems in correlation to a low strength, F specimens still exhibited a better thermal stability at 800º C than the other due to high amount of thermally-stable mullite phase (see Fig. 9). This could be attributed to the low-calcium AAM system of F samples, which possessed a highly polymerised 3D gel network as well as low content of bound water and water-rich Portlandite phase. This observation was also corroborated by the study of Kashani et al. [72], who concluded that the decline of Si/Al ratio associated with lower strength at elevated temperatures but superior thermal stability and a lower degree of cracking. The highest Ca/Si ratio (0.8 – 1.1) was recorded in CaO-rich S samples due to a very high amount of rigid C-A-S-H phase. However, the decomposition of C-A-S-H phase triggered severe cracks (see Fig. 11) and a significant drop in strength (see Fig. 4) after high temperature exposure. Moreover, a low Na/Al ratio reportedly correlated to a high intensity and depth of cracks [72]. It was also observed here where the Na/Al and Na/Si ratios seemed to reduce with increasing temperatures, which associated with the intensive cracking at 800ºC.

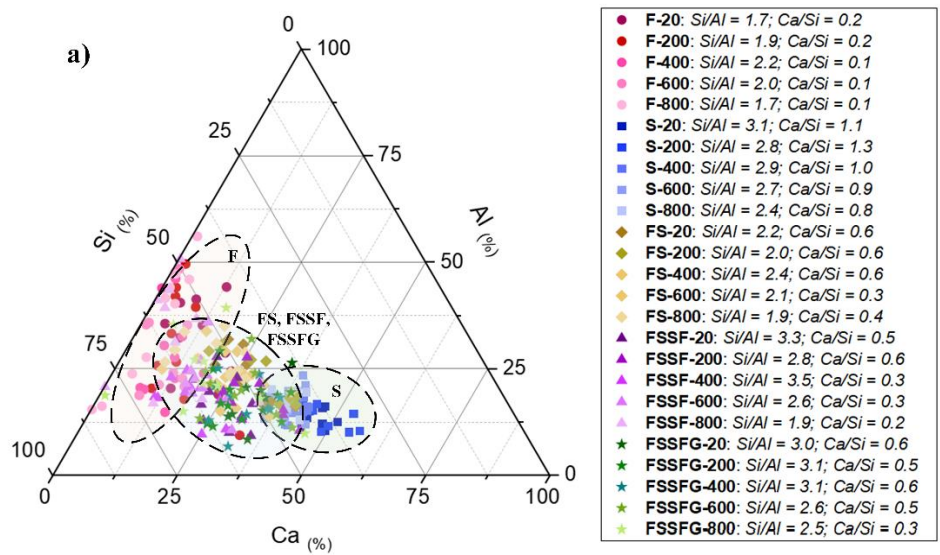

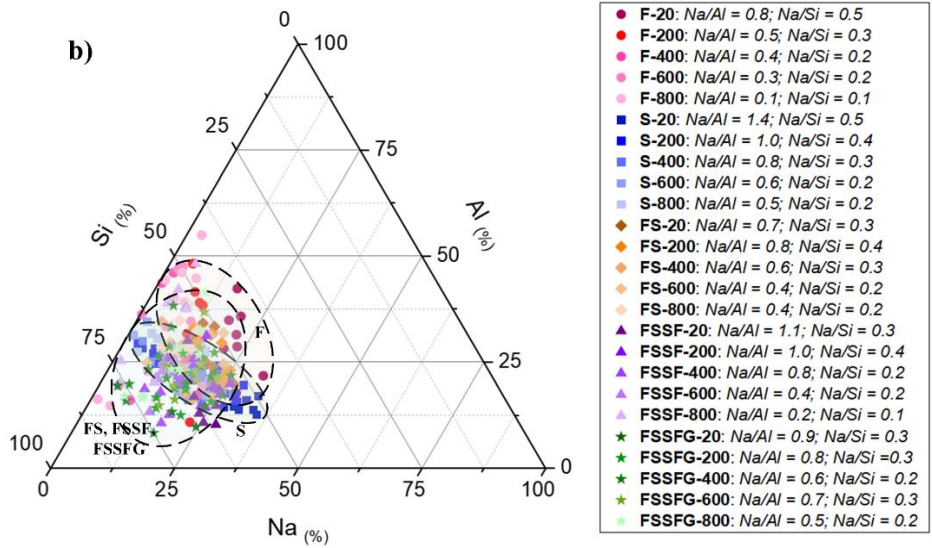

Fig. 12. Atomic plots obtained from EDS data corresponding to AAM mixes at elevated temperatures: a) Ca-Al-Si and b) Na-Al-Si ternary diagram.

**4.9 X-ray μCT**

The relative pore diameter and porosity evolution of AAMs at various heating temperatures is displayed in Table 3. Notably, the substantial changes in the microstructure of AAMs occurred between 600 and 800 °C, corresponding to the dehydration of C-(A)-S-H phase and crystallisation of binder gels (see Fig. 9). This phase transformation was reflected by an upsurge in total porosity of AAM samples after being subjected to 800°C, which led to the reduction in strength (see Fig. 4). This porous structure was also associated with the presence of nepheline [78], formed by exposure to 800 °C. Interestingly, FS samples achieved the least porous microstructure over other AAM mixtures within 20 – 600°C. Although the presence of GNPs slightly densified the microstructure of FSSFG – compared to FSSF samples, these two mixes with SF still showed higher porosity than FS specimens without SF usage. Less pore-filling effect of SF could be a reason for this observation due to a larger particle size of SF compared to FA particle used in this study (see Fig. 1).

Table 3. Pore characteristics of samples at different temperatures

| Sample ID | Temperature exposure | | | | | | | | | |
|---|---|---|---|---|---|---|---|---|---|---|
| | 20°C | | 200°C | | 400°C | | 600°C | | 800°C | |
| | $\phi_{avg}$ (mm) | $\varphi_t$ (%) | $\phi_{avg}$ (mm) | $\varphi_t$ (%) | $\phi_{avg}$ (mm) | $\varphi_t$ (%) | $\phi_{avg}$ (mm) | $\varphi_t$ (%) | $\phi_{avg}$ (mm) | $\varphi_t$ (%) |
| F | 0.14 | 1.24 | 0.12 | 1.39 | 0.09 | 4.89 | 0.1 | 5.9 | 0.11 | 11.27 |
| S | 0.18 | 1.18 | 0.14 | 1.04 | 0.23 | 2.91 | 0.23 | 4.37 | 0.22 | 9.43 |
| FS | 0.17 | 0.6 | 0.14 | 0.76 | 0.09 | 0.89 | 0.1 | 1.48 | 0.16 | 12.11 |
| FSSF | 0.15 | 1.74 | 0.12 | 2.49 | 0.12 | 2.82 | 0.11 | 3.11 | 0.12 | 12.46 |
| FSSFG | 0.09 | 1.62 | 0.11 | 2.12 | 0.12 | 2.5 | 0.12 | 2.8 | 0.13 | 13.76 |

$\phi_{avg}$: *Average diameter of pore;* $\varphi_t$: *Total porosity*

To visualise and compare the defects/damage of samples at elevated temperatures, 3D-rendered images and the volume fraction of defects (i.e., total porosity) – including cracks, close and interconnected pores within cross-sectional slices along the z-axis of samples, are estimated and presented in Fig. 13. As observed, the porosity distribution varied differently along the height of samples and increased with an upward temperature exposure. The shrinkage of AAM matrix was detected from 600°C, meanwhile the viscous sintering obviously occurred from 800°C with the formation of new crystalline phases. Apparently, there was a noticeable shift in pore structure at 800°C, especially in F and S samples, due to the increasingly porous matrix and severe crack propagation towards the surface induced by thermal shrinkage deformation and changes in the phase assemblage (as discussed in section 4.7). Only FS samples retained a relatively stable pore structure up to 600°C without significant changes in porosity. The incorporation of SF in both FSSF and FSSFG samples rendered the matrix slightly susceptible to elevated temperatures with more cracks and porosity. This was consistent with the thermal instability of the gel system with high Si/Al ratio, as discussed in section 4.8. For FSSFG specimens with GNPs, porosity distribution seemed to vary differently at several local areas after the exposure over 400°C. The densification facilitated by the high concentration of GNPs in several regions might not resist the thermal stress at elevated temperatures, resulting in the local cracks as peaks record (Fig. 13).

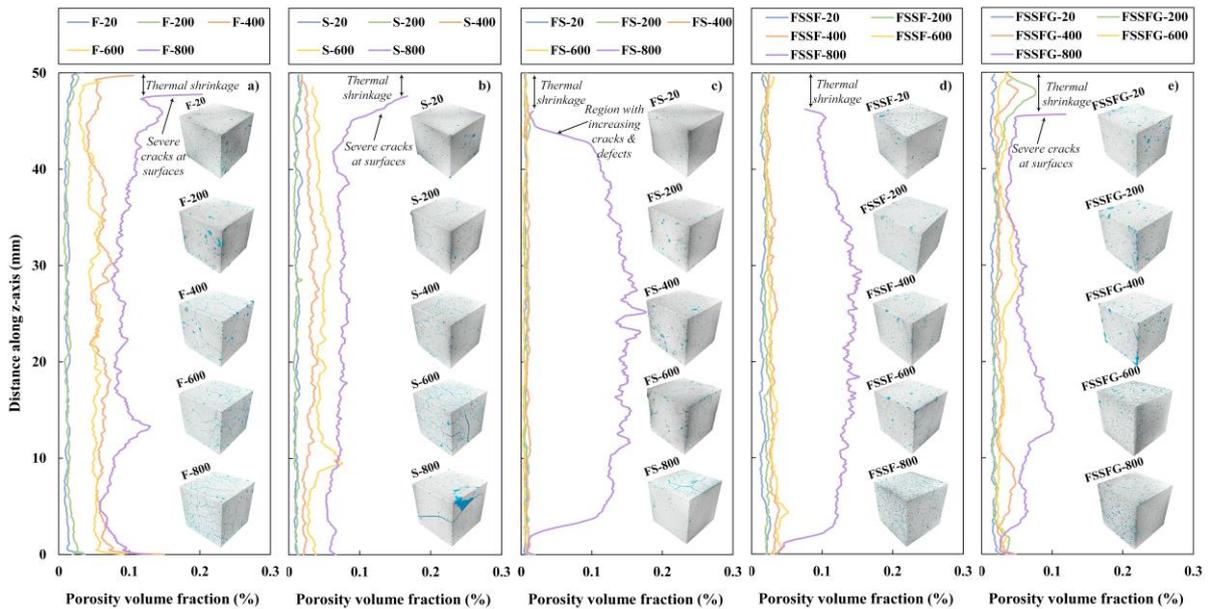

Fig. 13. Defect/porosity distribution along z-axis (cubic's height) at elevated temperatures corresponding to different AAM mixes: a) F samples, b) S samples, c) FS samples, d) FSSF samples, and e) FSSFG samples

Furthermore, it was observed that AAM samples obtained a relatively similar crack pattern after 800°C exposure. To analyse the crack propagation at elevated temperatures, 3D images were rendered and clipped to different depths from the top surface, as displayed in Fig. 14. Interestingly, due to the restrained conditions and the farthest distance to the surface (i.e., at 25-mm depth from surface), the thermal stress and built-up of vapour pressure were substantially amplified especially at the centre of specimens. This phenomenon increased the likelihood of cracks in the central area with a convergence of cracks at the core point of specimens (see Fig. 14). Subsequently, those central cracks branched and merged with adjacent defects/pores to form the densely interconnected channels towards the surface to release the vapour pressure. It should be noted that crack propagation was prone to follow pathways with the least constraint. Apparently, when moving towards the surface with less restraints (i.e., 25 → 10 → 5 mm depth from the surface), the crack branching transformed from dense and narrowly spaced to wider spacing with a larger crack width. Crack propagation was attributed to the devitrification and volume deformation of gels/phases (i.e., incongruent thermal shrinkage and expansion). The structural densification and viscous sintering softened the gels surrounding the pores, resulting in the shrinkage and cracking [72, 106]. In addition, the oxidation of $Fe_2O_3$-containing phases at elevated temperatures, especially in remnant unreacted FA (see. Fig. 11), could trigger the expansion cracking at the interfacial region between iron-rich spots and gels. The crystalline quartz phases might also induce the expansion cracking at high temperatures [107]. Those cracking mechanisms contributed to the significant reduction in load bearing capacity of AAM system and thus diminished its strength considerably at 800°C (see Fig. 4).

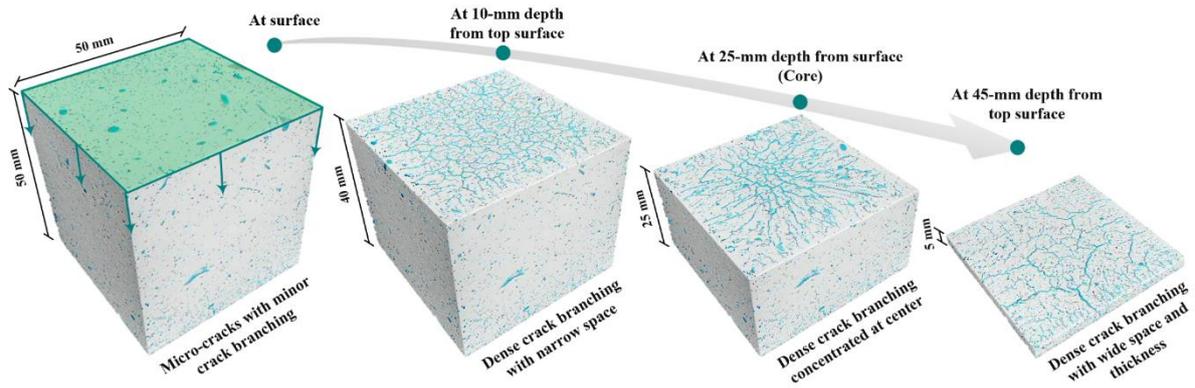

Fig. 14. Visualised crack patterns of representative AAM at different depth after 800°C exposure using X-ray μCT

## 5. Conclusions and recommendations

In this work, an ambient-cured one-part AAM was developed for high thermal performance at different elevated temperatures (200°C, 400°C, 600°C and 800°C). Five AAM mixes were compared and examined including F (100% FA), S (100% GGBFS), FS (60% FA + 40% GGBFS), FSSF (55% FA + 40% GGBFS + 5% SF) and FSSFG mixes (55% FA + 40% GGBFS + 5% SF + 0.1% GNPs). Based on empirical results, some key findings of this research can be summarised as follows:

1. Without heat energy consumption for curing, the nano-engineered ambient-cured AAM developed in this study still achieve high strength characteristics of approx. 80 MPa at 28 days. When exposing to elevated temperatures, high heat and build-up vapour somewhat establish an autoclaving condition inside AAM pastes, which promotes further geopolymerisation and hydration. This phenomenon facilitates the strength enhancement upon exposure to 400°C, where the FSSFG obtains a 28-day strength of approx. 100 MPa, followed by the FSSF and FS.
2. The higher the temperature exposure reached, the longer the heat energy can be stored by AAM. Specific heat capacity ($C_p$) of AAM, measured by Hot Disk, tends to decline upon elevated temperature exposure. The FSSFG paste can achieve relatively high $C_p$ value of approx. 1400 J/Kg°C at ambient condition.
3. The presence of GNPs in AAM matrices provides nucleating points for forming N-A-S-H and C-(A)-S-H bridges, thereby promoting the geopolymerisation/hydration reaction and crystallisation of amorphous phases. Increasing temperature exposure proportionally correlates to a higher fraction of crystalline phases formed.

4. Except for Ca-rich phases such as C-(A)-S-H and calcite, other phases in AAM matrices are highly stable with negligible phase change, corresponding to elevated temperature exposure up to 600ºC. The sintering process initiates around 800ºC, where structural instability of aluminosilicate gels triggers the formation of typical crystalline phases such gehlenite, wollastonite, akermanite and nepheline. The presence of crystalline akermanite and nepheline are highly associated with porous microstructure.
5. High Si/Al and Ca/Si ratios associated with high strength but low thermal stability. Low Na/Al and Na/Si ratios also correlate to intensive cracking.
6. Thermal expansion, shrinkage-induced stress and built-up of vapour pressure triggered intensive cracking in the core of specimen where it has the highest constraint effect. The dissipation of internal stresses led to crack propagation towards the surface with a shift in crack profile from narrow crack width with dense spacing to large crack width with wide spacing.
7. There is a promising use of nano-engineered one-part AAM as high-temperature binder, which can later on combine with aggregate for TES applications at around 400 – 600ºC.

Despite the good thermal performance of ambient-cured one-part AAM developed in this study, the matrix still contains a significant proportion of unreacted particles (see section 4.8). This is considered as one of the limitations of one-part AAM, which necessitates further research on the optimisation of mixing methods to maximise their hardened properties. The utilisation of microwave radiation curing can also be considered to improve microstructure. It should be noted that this study merely focused on high-temperature performance of AAM binder, which will be further developed in a combination with aggregate in the next study together with cyclic heat treatment to examine their potential for TES applications.

**CRediT authorship contribution statement**

**Nghia P. Tran**: Conceptualization, Methodology, Software, Data curation, Formal analysis, Investigation, Visualization, Validation, Writing – original draft. **Tuan N. Nguyen**: Methodology, Software, Data curation, Supervision, Writing – review & editing. **Jay R. Black**: Methodology, Data curation, Software, Writing – review & editing. **Tuan D. Ngo**: Methodology, Supervision, Writing – review & editing, Funding acquisition.

## Acknowledgement

This research is supported by the Building 4.0 CRC and Melbourne Research Scholarship for doctoral degree. The first author gratefully acknowledges the assistance of Vithushanthini Arulkumar and technical officers at the University of Melbourne, as well as the support from Dr. Rajeev Roychand and Dr. Yulin Patrisia at RMIT University. Thanks to the Melbourne Trace Analysis for Chemical, Earth and Environmental Sciences (TrACEES) Platform for access to the micro-CT scanner.